\documentstyle[12pt]{article}
\begin{document}
\thispagestyle{empty}
\newcommand{\be}{\begin{equation}}
\newcommand{\ee}{\end{equation}}
\newcommand{\sect}[1]{\setcounter{equation}{0}\section{#1}}
\newcommand{\vs}[1]{\rule[- #1 mm]{0mm}{#1 mm}}
\newcommand{\hs}[1]{\hspace{#1mm}}
\newcommand{\mb}[1]{\hs{5}\mbox{#1}\hs{5}}
\newcommand{\bea}{\begin{eqnarray}}
\newcommand{\eea}{\end{eqnarray}}
\newcommand{\wt}[1]{\widetilde{#1}}
\newcommand{\ux}[1]{\underline{#1}}
\newcommand{\ov}[1]{\overline{#1}}
\newcommand{\sm}[2]{\frac{\mbox{\footnotesize #1}\vs{-2}}
           {\vs{-2}\mbox{\footnotesize #2}}}
\newcommand{\prt}{\partial}
\newcommand{\eps}{\epsilon}\newcommand{\p}[1]{(\ref{#1})}
\newcommand{\R}{\mbox{\rule{0.2mm}{2.8mm}\hspace{-1.5mm} R}}
\newcommand{\Z}{Z\hspace{-2mm}Z}
\newcommand{\cd}{{\cal D}}
\newcommand{\cg}{{\cal G}}
\newcommand{\ck}{{\cal K}}
\newcommand{\cw}{{\cal W}}
\newcommand{\vj}{\vec{J}}
\newcommand{\vl}{\vec{\lambda}}
\newcommand{\vz}{\vec{\sigma}}
\newcommand{\vt}{\vec{\tau}}
\newcommand{\poiss}{\stackrel{\otimes}{,}}
\newcommand{\tx}{\theta_{12}}
\newcommand{\tb}{\overline{\theta}_{12}}
\newcommand{\zw}{{1\over z_{12}}}
\newcommand{\sqp}{{(1 + i\sqrt{3})\over 2}}
\newcommand{\sqm}{{(1 - i\sqrt{3})\over 2}}
\newcommand{\NP}[1]{Nucl.\ Phys.\ {\bf #1}}
\newcommand{\PLB}[1]{Phys.\ Lett.\ {B \bf #1}}
\newcommand{\PLA}[1]{Phys.\ Lett.\ {A \bf #1}}
\newcommand{\NC}[1]{Nuovo Cimento {\bf #1}}
\newcommand{\CMP}[1]{Commun.\ Math.\ Phys.\ {\bf #1}}
\newcommand{\PR}[1]{Phys.\ Rev.\ {\bf #1}}
\newcommand{\PRL}[1]{Phys.\ Rev.\ Lett.\ {\bf #1}}
\newcommand{\MPL}[1]{Mod.\ Phys.\ Lett.\ {\bf #1}}
\newcommand{\BLMS}[1]{Bull.\ London Math.\ Soc.\ {\bf #1}}
\newcommand{\IJMP}[1]{Int.\ J.\ Mod.\ Phys.\ {\bf #1}}
\newcommand{\JMP}[1]{Jour.\ Math.\ Phys.\ {\bf #1}}
\newcommand{\LMP}[1]{Lett.\ Math.\ Phys.\ {\bf #1}}
\newpage
\setcounter{page}{0} \thispagestyle{empty} \vs{12}

\renewcommand{\thefootnote}{\fnsymbol{footnote}}
{\par IFTUWr-03/2002/01 \par CBPF-NF-001/02}

\vspace{.5cm}
\begin{center}
{\large\bf Generalized Space-time Supersymmetries,\\ Division
Algebras and Octonionic M-theory} \vspace{.5cm}\\

Jerzy Lukierski ${}^a$\footnote{Supported by KBN grant 5P03B05620
}\footnote{{\em e-mail: lukier@ift.uni.wroc.pl}} and Francesco
Toppan ${}^b$\footnote{{\em e-mail: toppan@cbpf.br}}
\vspace{.1cm}\\

${}^a${\it Institute for Theoretical Physics, University of Wroc\l
aw,}
\\ {\it 50-204 Wroc\l aw, pl. Maxa Borna 9, Poland}\\
{\quad}\\ ${}^b${\it CBPF, CCP, Rua Dr. Xavier Sigaud 150,\\ cep
22290-180 Rio de Janeiro (RJ), Brazil}\vspace{.5cm}\\

{\bf Abstract}

\end{center}

We describe the set of generalized Poincar\'{e} and conformal
superalgebras in $D=4,5$ and $7$ dimensions as two sequences of
superalgebraic structures, taking values in the division algebras
${\bf R},{\bf C}$ and ${\bf H}$. The generalized conformal
superalgebras are described for $D=4$ by $OSp(1;8|{\bf R})$, for
$D=5$ by $SU(4,4;1)$ and for $D=7$ by $U_\alpha U(8;1|{\bf H})$.
The relation with other schemes, in particular the framework of
conformal spin (super)algebras and Jordan (super)algebras is
discussed. By extending the division-algebra-valued superalgebras
to octonions we get in $D=11$ an octonionic generalized
Poincar\'{e} superalgebra, which we call {\em octonionic
M-algebra}, describing the octonionic M-theory. It contains 32
real supercharges but, due to the octonionic structure, only $52$
real bosonic generators remain independent in place of the $528$
bosonic charges of standard $M$-algebra. In octonionic M-theory
there is a sort of equivalence between the octonionic M2
(supermembrane) and the octonionic M5 (super-$5$-brane) sectors.
We also define the octonionic generalized conformal
M-superalgebra, with $239$ bosonic generators.

\vfill \setcounter{page}0
\renewcommand{\thefootnote}{\arabic{footnote}}
\setcounter{footnote}0
\newpage
\section{Introduction}

We shall call generalized space-time supersymmetries the ones
which go beyond the standard H\L S scheme \cite{HLS}. In four
dimensions using the framework of local field theory and the
arguments from S-matrix theory it was shown \cite{{HLS}, {Soh}}
that the bosonic sector $B$ of Poincar\'{e} or conformal
superalgebra has the following tensor product
structure\footnote{Such a definition can be used also for standard
super-de Sitter algebras (for both signs of the radii).}:
\begin{eqnarray}
B &=& B_{geom}\oplus B_{int}, \label{Bdec}
\end{eqnarray}
where $B_{geom}$ describes space-time Poincar\'{e} or conformal
algebras and the remaining generators spanning $B_{int}$ are
Lorentz scalars. It is easy to show that one can introduce the
standard Poincar\'{e} superalgebra, satisfying the relation
(\ref{Bdec}), in any dimension (see e.g. \cite{Str}), but one
arrives at a difficulty in trying to impose in any dimension
standard conformal superalgebras. It appears \cite{{Nah},
{KT},{HLM}} that one can introduce only at $D=3,4$ and $6$ an
infinite series of standard conformal superalgebras, which can be
denoted in a unified way as $U_\alpha U(4;n|{\bf
F})$\footnote{$U_\alpha(n|{\bf F})$ describes the antiunitary
${\bf F}$-valued matrix group of transformations preserving the
${\bf F}$-valued antiHermitian bilinear form
${{q_i}^\dagger}A_{ij}q_j=inv$ ($A_{ij}=-{A_{ji}}^\dagger$) where
$q_i\in {\bf F}$ and $q_i\mapsto {q_i}^\dagger$ is the main
conjugation in ${\bf F}$. For quaternions (${\bf F}={\bf H}$) one
can show that $U_\alpha(n|{\bf H}) =O(2n; {\bf C})\cap
U(n,n)=O^\ast (2n)$. The supergroup $U_\alpha U(n;m|{\bf F})$
describes the ${\bf F}$-valued matrix supergroup of graded
transformations preserving the ${\bf F}$-valued bilinear form
${{q_i}^\dagger}A_{ij}q_j+{{\theta_k}^\dagger}\theta_k=inv$, where
$\theta_k$ ($k=1,\dots , m$) are ${\bf F}$-valued Grassmann
variables. For ${\bf F}={\bf H}$ one gets $U_\alpha U (n;m|{\bf
H})=SU(n,n;m)\cap OSp(2m,2n|{\bf C})$, which is usually denoted as
$OSp^\ast(2n;2m)$.} \cite{{HLM},{LN}} (${\bf F}={\bf R}$ for
$D=3$, ${\bf F}={\bf C}$ for $D=4$ and ${\bf F}={\bf H}$ for
$D=6$). More explicitly\footnote{We add here for completeness that
for $D=5$ there is a unique ``exotic" standard conformal
superalgebra $F_4$, with bosonic sector ${\overline{
O(5,2)}}\times SU(2)$ (see e.g. \cite{{BG},{DFV}}).}
\begin{eqnarray}
D=3:\quad U_\alpha U(4;n|{\bf R}) &\equiv& OSp(n;4|{\bf
R}),\nonumber\\ D=4:\quad U_\alpha U(4;n|{\bf C}) &\equiv&
SU(2,2;n), \label{d3d4}
\cr D=6:\quad U_\alpha U (4;n|{\bf H})  & \equiv & O^\ast Sp(8;
2n). \nonumber
\end{eqnarray}
It appears that if we wish to use the notion of conformal
superalgebra in any dimension we should extend the standard
Poincar\'{e} superalgebra (see e.g.
\cite{{AGIT},{Gun},{SorT},{Tow},{Ferr}}). The best known case is
in $D=11$, where the  generalized Poincar\'{e} superalgebra going
beyond the H\L S axioms is called the M-algebra and looks as
follows ($r,s= 1,2,\ldots , 32$; $\mu,\nu =0,1,\ldots , 10$):
\begin{eqnarray}
\relax &&\{ Q_r, Q_s\} = Z_{rs} = (C\Gamma_\mu)_{rs} P^\mu
+(C\Gamma_{[\mu\nu]})_{rs}Z^{[\mu\nu]}+
(C\Gamma_{[\mu_1\ldots\mu_5]})_{rs}Z^{[\mu_1,\ldots,\mu_5]},
\label{Malg}
\end{eqnarray}
where $C=\Gamma_0$ is the $D=11$ real Majorana charge conjugation
matrix. The generalized $D=11$ conformal superalgebra is obtained
by adding a second copy of the superalgebra (\ref{Malg}), with the
extension of the conformal accelerations sector to the $32\times
32$ symmetric matrices ${\tilde Z}_{rs}$:
\begin{eqnarray}
\relax \{S_r, S_s\} ={\tilde Z}_{rs} = (C\Gamma_\mu)_{rs} K^\mu
+(C\Gamma_{[\mu\nu]})_{rs}{\tilde Z}^{[\mu\nu]}+
(C\Gamma_{[\mu_1\ldots\mu_5]}){\tilde Z}^{[\mu_1,\ldots,\mu_5]}.
\end{eqnarray}
Both sets of generators $Z_{rs}$, ${\tilde Z}_{rs}$ are Abelian,
i.e.
\begin{eqnarray}
\relax && [Z_{rs}, Z_{tk}]=[{\tilde Z}_{rs}, {\tilde Z}_{tk}] =0.
\end{eqnarray}
It appears that if we introduce the crossed anticommutator,
completing the superalgebra relations
\begin{eqnarray}
\{ Q_r, S_s\} &=& L_{rs},
\end{eqnarray}
we get from the Jacobi identity that the $1024$ generators
$L_{rs}$ form the $GL(32;{\bf R})$ algebra \cite{Wes}.
Summarizing, the resulting superalgebra admits the following
five-grading
\begin{eqnarray}
&&\begin{array}{ccccc}
 I_{-2}& I_{-1} & I_0 & I_1 & I_2\\
  {\tilde Z}_{rs} &  S_r & L_{rs} & Q_r & Z_{rs}.
\end{array}
\label{5grad}
\end{eqnarray}
The set of generators $Z_{rs}$, ${\tilde Z}_{rs}$, $ L_{rs}$
describe the generalized $D=11$ conformal algebra $Sp(64)$
 (conformal M-algebra) and all the generators from (\ref{5grad})
form the superalgebra $OSp(1|64)$\cite{{vHvP},{Bar},{ABL},{Vas}} ,
known as generalized $D=11$ superconformal algebra (conformal
M-superalgebra).\par

The aim of this paper is to propose an analogous construction for
the sequence of  ${\bf F}$-valued (${\bf F}$=${\bf R}$, ${\bf C}$,
${\bf H}$, ${\bf O}$) generalized superalgebras, with the real
superalgebras describing generalized supersymmetries in $D=4$. We
shall describe these superalgebras in some detail in Sect. {\bf 2}
for $D=4$ ({\bf F}$=${\bf R}), $5$ ({\bf F}$=${\bf C}) and $7$
({\bf F}$=${\bf H}). We obtain the generalized Poincar\'{e}
superalgebras with $10$ real bosonic generators for $D=4$, $16$
real bosonic generators for $D=5$ and $28$ real bosonic generators
for $D=7$ and the corresponding $D=4,$ $5$ and $7$ generalized
conformal superalgebras $U_\alpha U(8;1|{\bf F})$.
\par
In Sect. {\bf 3} we shall consider the relation of our proposal to
other ways of introducing generalized supersymmetries, in
particular based on Lorentz spin and conformal spin algebras
\cite{{Fer},{AFLV},{AFL}}. It appears that our scheme for $D=7$
can be identified with the one following from the minimal
conformal spin algebra, but this is not the case for $D=4,5$. On
the other hand our generalized superalgebras can be called minimal
in another sense since the symmetrized product of supercharges
(i.e. the anticommutators) is spanned by the fundamental
representation of the respective Clifford algebras (${\bf
R}^4\times {\bf R}^4$ for $D=4$, ${\bf C}^4\times{\bf C}^4$ for
$D=5$, ${\bf H}^4\times {\bf H}^4$ for $D=7$)\footnote{The
fundamental representation of the Clifford algebra is its faithful
representation with minimal dimension \cite{{ABS},{Coq}}.}. The
proposal is linked with the generalized conformal and
superconformal algebra description in terms of ${\bf F}$-valued
Jordan (super)algebras \cite{{Kan},{Ced},{Gun}}.

In Sect. {\bf 4} we shall conjecture that one can use the proposed
superalgebras with the division algebra {\bf F} given by the
octonionic algebra ${\bf O}$\footnote{For the extension of
$U_\alpha U (n; m|{\bf F})$ algebra to octonions see also
\cite{{Sud},{HL}}.}. In particular we obtain in place of the
``standard" M-algebra (\ref{Malg}) an algebra which we call the
{\em octonionic M-algebra} with $52$ real bosonic generators
described by a $4\times 4$ octonionic Hermitian matrix. We provide
two alternative descriptions of the octonionic M-algebra: the
first one linear and bilinear in the octonionic $\Gamma$-matrices
and the second with their five-linear products only. We shall also
introduce an octonionic conformal M-superalgebra with $232$ real
bosonic generators.

In Sect. {\bf 5} we present the final remarks. In particular we
list some aspects of our framework which are postponed to further
consideration.

\section{The generalized $D=4,5$ and $7$ supersymmetries described
by {\bf F}-valued superalgebras ({\bf F} $=$ {\bf R}, {\bf C},
{\bf H})}

{\em i}) {\em Generalized Poincar\'{e} superalgebras.}\quad\par

The standard $N=1$ $D=4$ Poincar\'{e} superalgebra has the
following complex Hermitian form ($A,B =1,2$):
\begin{eqnarray}
\{ Q_A, {\overline Q}_{\dot{B}} \} &=& (\sigma_\mu)_{A\dot{B}}
P^\mu,\nonumber\\ \{Q_A, Q_B\} =0 && \{ {\overline Q}_{\dot A},
{\overline Q}_{\dot B}\} =0, \label{spa}
\end{eqnarray}
where $\sigma_\mu =( {\bf 1}_2, \sigma _i)$ describes the linear
basis of Hermitian $2\times 2$ matrices and $Q_A\mapsto
{Q_A}^\dagger = {\overline Q}^{\dot A}$ is the complex Hermitian
conjugation. One can however also introduce complex holomorphic
$D=4$ algebra as follows
\begin{eqnarray}
\{Q_A, {\overline Q}_{\dot B}\}&=&0,\nonumber\\
\{Q_A,Q_B\}=Z_{AB},&\quad& \{{\overline Q}_{\dot A}, {\overline
Q}_{\dot B}\} = {\overline Z}_{{\dot A}{\dot B}}.
\end{eqnarray}
Such types of superalgebras are the standard ones in the
description of $D=4$ Euclidean supersymmetry (see e.g.
\cite{{LN2},{NW}}). If we introduce the antisymmetric real
two-tensor field
\begin{eqnarray}
\relax Z_{[\mu\nu]} &=&\frac{1}{2i}({\sigma_{[\mu\nu]}}^{AB}Z_{AB}
-{{\tilde \sigma}_{[\mu\nu]}}^{{\dot A}{\dot B}} {\overline
Z}_{{\dot A}{\dot B}}),
\end{eqnarray}
one can incorporate both Abelian charges $P^\mu$, $Z^{[\mu\nu]}$
in the Majorana form of $D=4$ superPoincar\'e algebra
($a,b=1,\ldots,4$, $\mu,\nu = 0,1,2,3$)
\begin{eqnarray}
\relax &&\{Q_a,Q_b\} = X_{ab}\equiv (C{\Gamma^{(4)}}_\mu
)_{ab}P^\mu + (C{\Gamma^{(4)}}_{[\mu\nu]})_{ab}Z^{[\mu\nu]}.
\label{ext}
\end{eqnarray}
We see that replacing in $D=4$ the complex-Hermitian structure by
a real one we find a place for six new Abelian tensorial charges.
Such superalgebra was recently considered \cite{{FP},{BL},{GGHT}}
as the $D=4$ counterpart of the M-algebra (\ref{Malg}) and
describes the supersymmetric theories with domain walls and
four-dimensional supermembranes. One can state that the physical
background for the extension (\ref{ext}) of the standard
Poincar\'e superalgebra (\ref{spa}) is now established. It is the
starting point of our construction. In order to describe the $D=5$
and $D=7$ extended Poincar\'e superalgebras we generalize
(\ref{ext}) to the {\bf F}-valued\footnote{${\bf F} ={\bf C}$ and
${\bf H}$; see however Sect. 4 where we consider ${\bf F}={\bf
O}$. We add that an abstract algebra $\{A,B\}=C$ of matrices
having entries ($a_{ij}$, $b_{ij}$, $c_{ij}$ respectively) valued
in a division algebra ${\bf F}$ (i.e. $a_{ij} =\sum_{\alpha}
{a_{ij}}_\alpha \tau_\alpha$, with $\tau_0$ the identity, and
similarly for $B$, $C$), implies the following relations on the
real components ${c_{ik}}_\gamma =\sum_{j,\alpha,\beta}
\{{a_{ij}}_\alpha , {b_{jk}}_\beta\} C_{\alpha\beta\gamma}$, where
$C_{\alpha\beta\gamma}$ are the structure constants of ${\bf F}$.}
Hermitian superalgebras
\begin{eqnarray}
\{Q_a, {Q^\dagger}_b\} = Z_{ ab}, &\quad& Z_{ab} =
{Z_{ba}}^\dagger,\nonumber\\ &\{Q_a, Q_b\} =0,& \label{Fspa}
\end{eqnarray}
where $\dagger$ denotes the principal conjugation in the {\bf
F}-algebra, namely \begin{eqnarray} {\bf C}: &&
\begin{array}{ll} Q_a=
{Q_a}^0+i{Q_a}^1,\quad& {Q_a}^\dagger
={Q_a}^0-i{Q_a}^1,\nonumber\\ Z_{ab}= {X_{ab}}^0+i{Y_{ab}}^1,\quad
& {Z_{ab}}^\dagger ={X_{ab}}^0-i{Y_{ab}}^1, \end{array}\nonumber\\
{\bf H}: &&
 \begin{array}{ll} Q_a=
{Q_a}^0+e_r{Q_a}^{(r)},\quad & {Q_a}^\dagger
={Q_a}^0-e_r{Q_a}^{(r)},\\ Z_{ab}=
{X_{ab}}^0+e_r{Y_{ab}}^{(r)},\quad &{Z_{ab}}^\dagger
={X_{ab}}^0-e_r{Y_{ab}}^{(r)} .
\end{array}
\end{eqnarray} In order to write
the superalgebra (\ref{Fspa}) in a Dirac matrices basis we shall
at first introduce $2\times 2$ complex Dirac matrices for the
$O(3)\simeq SU(2)$ algebra (i.e. the three Pauli matrices
$\sigma_r$, $r=1,2,3$) and $2\times 2$ quaternionic matrices for
the $O(5)\simeq SU(2;{\bf H})$ algebra
\begin{eqnarray}
O(5):&& \Sigma_r=\left(\begin{array}{cc}
  0 & e_r \\
  -e_r & 0
\end{array}\right),\quad \Sigma_4=\left(\begin{array}{cc}
  1 & 0 \\
  0 & -1
\end{array}\right),\quad
\Sigma_5=\left(\begin{array}{cc}
  0 & 1\\
  1& 0
\end{array}\right),
\label{05}\end{eqnarray} where $e_r$ are the three quaternionic
units. Then we shall consider $O(4,1)$ and $O(6,1)$ as $D=3$ and
respectively $D=5$ Euclidean conformal algebras and follow the
rules (see e.g. \cite{Oku}) in order to introduce Dirac's $\gamma$
matrices for $O(p+1,q+1)$ by doubling the dimension of the
$O(p,q)$ representations. We obtain

{\em i1}) $D=5$.

\begin{eqnarray}
&& {\Gamma_r}^{(5)} =\left(\begin{array}{cc}
  0 & \sigma_r\\
  \sigma_r& 0
\end{array}\right),\quad
{\Gamma_4}^{(5)} =\left(\begin{array}{cc}
  {\bf 1}_2 & 0\\
  0& -{\bf 1}_2
\end{array}\right),\quad
{\Gamma_0}^{(5)} =\left(\begin{array}{cc}
  0 & {\bf 1}_2\\
  -{\bf 1}_2& 0
\end{array}\right).
\label{gammas}
\end{eqnarray}
One can describe the complex Hermitian $4\times 4$ matrices as
linear combination of $16$ Hermitian-symmetric matrices
${\Gamma_\mu}^{(5)}C^{(5)}$, $ {\Gamma_{\mu\nu}}^{(5)}C^{(5)}$ and
$iC^{(5)}$. One sets
\begin{eqnarray}
\relax &&\{Q_a,{Q_b}^\dagger\} = Z_{ab}=
({\Gamma_\mu}^{(5)}C^{(5)})_{ab}
P^\mu+({\Gamma_{[\mu\nu]}}^{(5)}C^{(5)})_{ab}Z^{[\mu\nu]} + i
{C^{(5)}}_{ab}Z, \label{cspa}
\end{eqnarray}
where $C^{(5)}$ is the $O(4,1)$ complex charge conjugation matrix
satisfying the relations
\begin{eqnarray}
{{\Gamma_\mu}^{(5)}}^\dagger
C^{(5)}&=&-C^{(5)}{{\Gamma_\mu}^{(5)}},\nonumber\\
{C^{(5)}}^\dagger &=& -C^{(5)}.\label{chcon} \end{eqnarray} In the
representation with ${\Gamma_a}^{(5)}={{\Gamma_a}^{(5)}}^\dagger$
($a=1,2,3,4$) (see e.g. (\ref{gammas})) and
${\Gamma_0}^{(5)}=-{{\Gamma_0}^{(5)}}^\dagger$, we should put
$C^{(5)} ={\Gamma_0}^{(5)}$. The maximal covariance algebra of the
supercharges is given by the group $GL(4,{\bf C})$, however
distinguished role is played by its subgroup $U_\alpha(4;{\bf
C})=U(2,2)$, because sixteen generators $Z_{ab}$ from (\ref{cspa})
belong to the adjoint representation of $U(2,2)$.

{\em i2}) $D=7$.

The Hermitian quaternionic representation of the $O(6,1)$ Clifford
algebra can be obtained from (\ref{gammas}) as follows
($p=1,\ldots,5$, $\mu=0,1,\ldots,6$)
\begin{eqnarray}
&& {\Gamma_p}^{(7)} =\left(\begin{array}{cc}
  0 & \Sigma_p\\
  \Sigma_p& 0
\end{array}\right),\quad
{\Gamma_6}^{(7)} =\left(\begin{array}{cc}
  {\bf 1}_2 & 0\\
  0& -{\bf 1}_2
\end{array}\right),\quad
{\Gamma_0}^{(7)} =\left(\begin{array}{cc}
  0 & {\bf 1}_2\\
  -{\bf 1}_2& 0
\end{array}\right).
\label{gammabis}
\end{eqnarray}
The $O(6,1)$ quaternionic charge conjugation matrix $C^{(7)}$
satisfies relations analogous to (\ref{chcon}) with
quaternionic-Hermitian conjugation. In the representation with
${\Gamma_k}^{(7)}={ {\Gamma_k}^{(7)}}^\dagger$ ($k=1,2,3,4,5,6$)
and ${\Gamma_0}^{(7)}= -{{\Gamma_0}^{(7)}}^\dagger$ (see e.g.
(\ref{gammabis})) we obtain again $C^{(7)} = {\Gamma_0}^{(7)}$.

If we consider the symmetry properties of the products
$C\Gamma_{[\mu_1,\ldots,\mu_k]}$ ($k=1,\ldots,7$) under
quaternionic conjugation we obtain that (\ref{Fspa}) for ${\bf
F}={\bf H}$ can be decomposed as follows ($\mu,\nu=0,1,\ldots,6$):
\begin{eqnarray}
&& \{Q_a,{Q_b}^\dagger\}= Z_{ab}=
(C^{(7)}{\Gamma_\mu}^{(7)})_{ab}P^\mu+
(C^{(7)}{\Gamma_{[\mu\nu]}}^{(7)})_{ab} Z^{[\mu\nu]}.
\label{decompo}
\end{eqnarray}
The most general covariance group of quaternionic Poincar\'e
algebra (\ref{decompo}) is $GL(4,{\bf H})$, and its distinguished
subgroup is $U_\alpha(4,{\bf H})\simeq SO^\ast(8)\simeq SO(6,2)$.
The $28$ bosonic real generators spanning $Z_{ab}$ in
(\ref{decompo}) are described by the adjoint representation of
$U_\alpha(4,{\bf H})$.

{\em ii}) {\em Generalized conformal superalgebras.}

Following the procedure of obtaining $OSp(1;64)$ from the
M-algebra (\ref{Malg}) one can add a second copy of the ${\bf
F}$-valued superalgebra (\ref{Fspa})
\begin{eqnarray}
&&\{S_a,S_b\} = {\tilde Z}_{ab}
\end{eqnarray}
and impose the Jacobi identities which imply that the mixed
anticommutator $\{Q_a,S_b\}=L_{ab}$ describes the $GL(4|{\bf F})$
Lie algebra generators. One obtains the following five-fold graded
structure
\begin{eqnarray}
&&\begin{array}{ccccc}
 I_{-2}& I_{-1} & I_0 & I_1 & I_2\\
  {\tilde Z}_{ab} &  S_a & L_{ab} & Q_a & Z_{ab}.
\label{gradgen}
\end{array}
\end{eqnarray}
If ${\bf F}={\bf R}$ the set of  generators (\ref{gradgen})
describe the $D=4$ generalized conformal superalgebra $OSp(1|8)$
with its bosonic sector describing the $D=4$ generalized conformal
algebra.

The construction for $D=5$ and $D=7$ corresponds respectively to
${\bf F}={\bf C}$ and ${\bf F}={\bf H}$.

{\em ii1}) $D=5$

In such a case the generalized conformal superalgebra is complex.
The complex generators $Z_{ab}$, ${\tilde Z}_{ab}$ in
(\ref{gradgen}) describe complex Hermitian algebras (see
(\ref{cspa})), and $L_{ab}$ span the $GL(4|{\bf C})$ algebra. It
can be checked that the complex bosonic algebra with three-grading
\begin{eqnarray}
&&\begin{array}{ccc}
 I_{-2}& I_0  & I_2\\
  {\tilde X}_{ab} &  L_{ab} & X_{ab}.
\label{gradthree}
\end{array}
\end{eqnarray}
describes the $U_\alpha(8,{\bf C})=U(4,4)$ algebra which is our
$D=5$ generalized conformal algebra. The five-grading
(\ref{gradgen}) provides $SU(4,4;1)$ as $D=5$ generalized
conformal superalgebra.

{\em ii2}) D=7

This case corresponds to inserting in (\ref{gradgen}) into the
sectors $I_2$ and $I_{-2}$ two copies of the $D=7$ Poincar\'e
superalgebra given by (\ref{decompo}). The sector $I_0$ is then
described by $GL(4;{\bf H})\simeq{SU^\ast (8)}$ algebra, and the
quaternionic three-graded algebra (\ref{gradthree}) provides
$U_\alpha (8|{\bf H})\simeq O^\ast(16)$ as the $D=7$ generalized
conformal algebra. The supersymmetric extension can be obtained by
imposing the five-grading (\ref{gradgen}) and it leads to the
$D=7$ generalized conformal superalgebra $U_\alpha U(8;1|{\bf
H})$.

Summarizing we see that the $D=4$, $D=5$ and $D=7$ generalized
conformal algebras and generalized conformal superalgebras are
given respectively by $U_\alpha (8|{\bf F})$ and $U_\alpha
U(8;1|{\bf F})$. We obtain the following numbers of additional
(in comparison with $O(D,2)$) 
bosonic generators, which are present in generalized conformal
algebras and conformal superalgebras:
\begin{eqnarray}
\begin{array}{ll|c}
   &\vline \quad\frac{U_\alpha(8;{\bf F})}{O(D,2)} &\quad
\frac{U_\alpha(8;{\bf F})\times   U(1;{\bf F})}{O(D,2)} \\ \hline
  D=4\quad ({\bf F}={\bf R}) &\vline\quad 21\quad (10) & \quad
21\quad  \\ \hline
  D=5\quad ({\bf F}={\bf C}) &\vline \quad 42\quad (16) & \quad
43\quad \\
  \hline
  D=7\quad ({\bf F}={\bf H}) &\vline \quad 84\quad (28) & \quad
87\quad 
\end{array}
\end{eqnarray}
where in brackets we provided the number of bosonic generators
which appear if we pass from the standard to the generalized
Poincar\'e superalgebra. These generators can also be treated as
introducing an extended $D$-dimensional space-time (see e.g.
\cite{{BL},{ABL},{Vas}}) with additional tensorial coordinates
besides the Minkowski space-time variables. We obtain
\begin{eqnarray}
\begin{tabular}{c|c|l}
   & 
     standard  space-time &  
   extended\quad space-time
    \\ \hline
  D=4 &
  4 &
\quad    4   \, + \, 6 = 10 \\ \hline
  D=5 &
    5 &
\quad     5 \, + \,  11 = 16\\
  \hline
  D=7  &
  7 &
\quad   7 \, + \,  21 = 28
\end{tabular}
\end{eqnarray}
The field realizations on space-time with additional coordinates
can be related with the representations of infinite-dimensional
spin algebras with infinite spin or helicity spectra \cite{Vas}.

\section{Relations with spinor algebras, representations
of Clifford algebras and Jordan algebras}

The existence of standard conformal supersymmetries at $D=3,4$ and
$6$ described by the set of superalgebras $U_\alpha U(4;n,{\bf
F})$ follows from  the property that the spinorial coverings of
the conformal algebra $O(D,2)$ are described for $D=3,4,6$ by
$U_\alpha(4|{\bf F})$, i.e.
\begin{eqnarray}
&&Spin(3,2)=\overline{O(3,2)}=Sp(4;{\bf R}),\nonumber\\
&&Spin(4,2)=\overline{O(4,2)}=\overline{SU(2,2)},\nonumber\\
&&Spin(6,2)=\overline{O(6,2)}=U_\alpha(4|{\bf H})=O^\ast(8,{\bf
C}).\label{spincov}
\end{eqnarray}
For $D=5$ and $D>6$ the spinorial covering of $\overline{O(D,2)}$
is not described by a classical Lie group.

Recently the notion of spin algebra has been introduced
\cite{{Fer},{AFLV},{AFL}} in any dimension generalizing the notion
of standard spin covering (\ref{spincov}). Let us firstly
introduce for the orthogonal group $O(n,m)$  its fundamental
spinorial representation described by an $N$-dimensional vector
space ${\bf F}_{(n,m)}$ with the choice of ${\bf F}$ (${\bf R}$,
${\bf C}$ or ${\bf H}$) depending on the pair of numbers
$(n,m)$\footnote{The fundamental spinor representation is
determined by the minimal faithful Clifford algebra representation
of $O(n,m)$ with generators $I_{\mu\nu} = \frac{1}{2}[\Gamma_\mu
,\Gamma_\nu]$.}. The spin group $Spin(n,m)$ is the ${\bf
F}$-valued $N\times N$ matrix Lie group of endomorphisms of ${\bf
F}_{(n,m)}$ which contains the spinorial covering of
$\overline{O(n,m)}$\footnote{In \cite{{Fer},{AFLV},{AFL}}
$Spin(n,m)$ are real algebras; we assume that $Spin(n,m)$ are
${\bf F}_{(n,m)}$-valued matrices. Both descriptions are
equivalent.}
\begin{eqnarray}
\overline{O(n,m)} \subset Spin(n,m), \quad D=n+m.
\end{eqnarray}
In particular one can distinguish the minimal spin group
$Spin_{min}(n,m)$, with minimal number of real generators. In the
standard case $D=3,4$ or $6$ we have $Spin_{min}
(D,2)=\overline{O(D,2)}$, but for $D=5$ and $D\geq 7$ we obtain
that $dim~ Spin_{min}(D,2)> dim~ O(D,2)=\frac{1}{2}(D+1)(D+2)$. If
we supersymmetrize the $Spin_{min}(D,2)$ algebra we obtain the
minimal conformal spin superalgebra
${\widetilde{Spin}}_{min}(D,2)$. In $D=3,4$ and $6$ one gets
${\widetilde{Spin}}_{min}(D,2)=U_\alpha U(4;n|{\bf H})$.

{\em i}) $D=5$.

It can be shown that for $D=5$ the fundamental conformal spinors
${\bf F}_{(4,1)}\equiv {\bf H}^4$ and (see
\cite{{Fer},{AFLV},{AFL}})
\begin{eqnarray}
&& Spin_{min}(5,2) =U_\alpha(4 ,{\bf H})\simeq O^\ast(8)\simeq
O(6,2),\label{spinmin}\\ &&{\widetilde{Spin}}_{min} (5,2)
=U_\alpha U(4;n|{\bf H}).\label{spinmin2}
\end{eqnarray}
The formula (\ref{spinmin}) describes the $D=6$ conformal algebra
and the relation (\ref{spinmin2}) assigns as minimal $D=5$ spin
superalgebra the standard $D=6$ conformal superalgebra (see
(\ref{d3d4})) with bosonic sector $O(6,2)\times O(3)$. In order to
interpret $U_\alpha (4)$ as $D=5$ conformal spin algebra with
$D=5$ tensor structure  we should perform the dimensional
reduction $D=6\mapsto D=5$. The seven generators spanning the
coset $\frac{O(6,2)}{O(5,2)}$ are described by $O(5,2)$
seven-vector, and after dimensional reduction they will form an
$O(4,1)$ five-vector and two $D=5$ scalars. These seven generators
will extend the $D=5$ conformal algebra $O(5,2)$. In the
supersymmetric case $U_\alpha U(4;n|{\bf H})$ is used as $D=5$
conformal spin superalgebra and will contain, besides the seven
generators from $\frac{O(6,2)}{O(5,2)}$, also three scalar $O(3)$
generators ($U(1|{\bf H})\simeq SU(2)\simeq O(3)$) describing the
internal symmetry sector\footnote{This internal sector is usually
referred to as describing $R$-symmetries.}.

From the considerations in Sect. 2 follows that
 our $D=5$ generalized conformal
algebra $SU(4,4)\supset U_\alpha(4;{\bf H})$. Subsequently, for
our $D=5$ generalized conformal superalgebra we obtain
\begin{eqnarray}
SU(4,4;2)&\supset& U_\alpha U(4;1|{\bf H}),
\end{eqnarray}
but $U_\alpha U(4;1|{\bf H})$  is not contained in $SU(4,4;1)$ in
analogy with the relations between $D=4$ standard and generalized
conformal superalgebras \cite{NPWH}, where $OSp(8;2)\supset
SU(2,2;1)$, but one cannot embed $SU(2,2;1)$ into $OSp(8;1)$.

It is easy to see that $SU(4,4)$ in comparison with minimal $D=5$
conformal spin algebra $SU(4,4)\supset U_\alpha(4;{\bf H})$
contains more additional bosonic generators. In fact the principle
of constructing our generalized supersymmetries in $D=4$ and $D=5$
are analogous. In $D=4$ we relaxed the restrictions on
superalgebra by replacing the complex structure by a real one, and
in $D=5$ the quaternionic structure is replaced by the complex
one.

{\em ii}) $D=7$.

In $D=7$ the situation is different. The fundamental $D=7$
conformal spinors are given by ${\bf H}^8$ and our generalized
conformal superalgebra $U_\alpha U (8;1,{\bf H})$ is identical
with the minimal conformal spin superalgebra
\begin{eqnarray}
Spin_{min} (7,2) &=& U_\alpha U(8;1|{\bf H})
\end{eqnarray}
with bosonic sector containing 123 bosonic generators (36 $O(7,2)$
generators $+$ 84 additional tensorial generators $+$ 3 generators
describing $U(1|{\bf H}) = SU(2)$ R-symmetry).

In order to compare the minimal conformal fundamental spin
algebras \cite{{Fer},{AFLV},{AFL}} with our generalized conformal
algebras let us write for $D=4,5$ and $7$ the fundamental Lorentz
spin representations and minimal Clifford algebra modules
permitting to represent faithfully $O(D-1,1)$ $\Gamma$-matrices:
\begin{eqnarray}
\begin{array}{lll}
   &\vline \quad min.\quad spin \quad F_{(D-1,1)} &\vline\quad
   min.\quad Clifford \quad mod. \quad C_{(D-1,1)}
    \\ \hline
  D=4 &\vline\quad\quad\quad\quad {\bf C}^2 &\vline \quad\quad\quad\quad
{\bf R}^4 \\ \hline
  D=5 &\vline \quad\quad\quad\quad {\bf H}^2 &\vline \quad\quad\quad\quad
{\bf C}^4\\
  \hline
  D=7\quad\quad\quad\quad &\vline \quad\quad\quad\quad {\bf H}^4 &\vline
\quad\quad\quad\quad
  {\bf H}^4
\end{array}
\label{table}
\end{eqnarray}
We see that our generalized Poincar\'e supercharges $Q_a$ are
described by minimal Clifford algebra modules $C_{(D-1,1)}$. We
also see from (\ref{table}) why there is a difference between the
conformal spin superalgebra approach with supercharges described
by $F_{(D-1,1)}$ and our proposal for $D=4$ and $D=5$.

We would like to mention here that the sequence of superalgebras
(12) as well as the corresponding conformal algebras (23) and
conformal superalgebras (22) can be put in the framework of
Jordanian algebras and Jordanian superalgebras [33,11]. Our goal
here was to assign concrete generalized supersymmetries to the
particular ${\bf F}$-valued chains of superalgebras. It should be
added, that in the framework of Jordanian (super)algebras one can
also include $3\times 3$ octonionic-Hermitian algebra of matrices
$J_3 ({\bf O})$, but the extesion of the generators $Z_{ab}$ to
octonionic-valued $4\times 4$ Hermitian matrices with 52 real
generators is not included into Jordan superalgebras sequence.
Such a superalgebra, with 32 real supercharges due to the imposed
octonionic-Hermitian structure will be called {\em octonionic
$M$-algebra}. In the next section we shall discuss our
octonionic-valued superalgebras.

\section{Octonionic M-superalgebras and Octonionic M-theory}

One of the features of the proposed sequence of generalized
supersymmetries is the possibility of extending the ${\bf
F}$-valued superalgebra structures to octonions.  Octonions are
described by eight real numbers ($k=1,\ldots ,7$)
\begin{eqnarray}
X\in {\bf O}: &\quad& X=X_0+X_k t_k,
\end{eqnarray}
where the seven octonionic units $t_k$ satisfy the nonassociative
algebra
\begin{eqnarray}
t_kt_l&=&-\delta_{kl}+\frac{1}{2}{f_{kl}}^mt_m.
\end{eqnarray}
The octonions are endowed with the principal involution
${\overline t}_k=-t_k$, and unit octonions describe the unit
sphere $S^7$ through $X{\overline X}=1$\footnote{One can also say
that $S^7$ describes the octonionic extension $U(1|{\bf O})$ of
$U(1)$; $S^7$ is however not a Lie group, but rather the so-called
soft Lie group \cite{Soh2}.}.

{\em i}) {\em The octonionic Poincar\'e M-superalgebra (octonionic
M-algebra).}

From extending (\ref{Fspa}) to octonions $^{\small 6}$ it follows
that in $D=11$ ($a=1,\ldots,4$)
\begin{eqnarray}
&\{Q_a, {Q_b}^\dagger\}= Z_{ab},&\nonumber\\ &\{Q_a, Q_b\}
=\{{Q_a}^\dagger,{Q_b}^\dagger\}=0,& \label{Moctalg}
\end{eqnarray}
where
\begin{eqnarray}
Q_a &=& {Q_a}^0+{Q_a}^{(k)}t_k,\nonumber\\ Z_{ab} &=&
{Z_{ab}}^0+{Z_{ab}}^{(k)}t_k
\end{eqnarray}
and $Z_{ab} = {Z_{ab}}^\dagger = $ implies that ${Z_{ab}}^0=
{Z_{ba}}^0$, ${Z_{ab}}^{(k)}=-{Z_{ba}}^{(k)}$, i.e. the algebra
(\ref{Moctalg}) is described by $52$  bosonic generators.
Following (\ref{05}) one can introduce the octonionic $2\times 2$
gamma matrices ($k=1,\ldots,7$) realizing the $(9,0)$
signature\footnote{Equivalently, to construct $D=11$ octonionic
gamma matrices we could start from the octonionic realization of
Clifford algebra with $(1,8)$ signature. It is worth mentioning
that octonionic realizations of Clifford algebras only exist in
specific signatures, such as $(0,7)$, $(9,0)$, $(1,8)$, $(10,1)$,
$(2,9)$ etc. They are related to the nonassociative realizations
of $D=1$ $N$-extended supersymmetries (see \cite{Top}), which are
beyond the classification of the representations of associative
$D=1$ $N$-extended supersymmetries \cite{PaTo}, based on the
Clifford algebras over ${\bf R}$, ${\bf C}$ and ${\bf H}$.}
\begin{eqnarray}
&& {\Sigma_k} =\left(\begin{array}{cc}
  0 & t_k\\
  -t_k& 0
\end{array}\right),\quad
{\Sigma_8} =\left(\begin{array}{cc}
  1 & 0\\
  0& -1
\end{array}\right),\quad
{\Sigma_9} =\left(\begin{array}{cc}
  0 & 1\\
  1 & 0
\end{array}\right),
\end{eqnarray}
and further the following $4\times 4$ $D=11$ octonionic
$\Gamma_\mu$ matrices ($R=1,\ldots ,9$)
\begin{eqnarray}
&& {\Gamma_R}^{(11)} =\left(\begin{array}{cc}
  0 & \Sigma_R\\
  \Sigma_R& 0
\end{array}\right),\quad
{\Gamma_{10}}^{(11)} =\left(\begin{array}{cc}
  {\bf 1}_2 & 0\\
  0& -{\bf 1}_2
\end{array}\right),\quad
{\Gamma_0}^{(11)} =\left(\begin{array}{cc}
  0 & {\bf 1}_2\\
  -{\bf 1}_2& 0
\end{array}\right),
\label{gamma11}
\end{eqnarray}
with the $D=11$ matrix $C^{(11)}$ given again by the matrix
${\Gamma_0}^{(11)}$ and satisfying relations analogous to
(\ref{chcon}) with octonionic principal conjugation. Taking into
consideration that ${\Gamma_\mu}^{(11)}$ for $\mu =1,\ldots, 10$
are octonionic-Hermitian and ${\Gamma_0}^{(11)}$ is antihermitian,
one can write (\ref{Moctalg}) as follows
\begin{eqnarray}
\{Q_a, {Q_b}^\dagger\}&=& ({C}^{(11)}{\Gamma_\mu}^{(11)})_{ab}
P^\mu +({C}^{(11)} {\Gamma_{\mu\nu}}^{(11)})_{ab} Z^{\mu\nu}.
\label{newsusy}
\end{eqnarray}
From the multiplication table of the octonions  follows that
($k,l=1,\ldots, 7$)
\begin{eqnarray}
\relax &&
{\Gamma_{kl}}^{(1)}={\Gamma_{[k}}^{(11)}{\Gamma_{l]}}^{(11)} =
{f_{kl}}^m {\Gamma_m}^{(11)}{\Gamma_8}^{(11)}
{\Gamma_9}^{(11)}{\Gamma_{10}}^{(11)}{\Gamma_{11}}^{(11)}.
\label{autom}
\end{eqnarray}
We see that out of the $21$ bilinear products of the first seven
matrices (\ref{gamma11}) only $7$ are independent and they
correspond to the generators of $\frac{O(7)}{G_2}$. The remaining
$4\times 4$ octonionic $\Gamma$-matrices in the antisymmetric
products $\relax {\Gamma_{\mu\nu}}^{(11)} = {\Gamma_{[\mu}}^{(11)}
{\Gamma_{\nu]}}^{(11)}$ are linearly independent, i.e. we get
$55-14=41$  generators describing the coset $\frac{O(10,1)}{G_2}$,
 which plays also the role of octonionic $D=11$ Lorentz algebra. It
 is easy to see that the maximal number of real generators on the
r.h.s. of (\ref{Moctalg}) is $52=11 +41$, i.e. the relation
(\ref{newsusy}) indeed saturates the octonionic-valued
anticommutator $\{Q_a,{Q_b}^\dagger \}$ ($a,b=1,2,3,4$).

It should also be stressed that in the definition of $n>2$
antisymmetric products of the octonionic $\Gamma$-matrices
(\ref{gamma11}) one should provide the order of multiplication,
because the Dirac algebra with the basis (\ref{gamma11}) is
non-associative. To be explicit, the antisymmetrized product of
$n$ octonionic matrices $A_i$ ($i=1,2,\dots, n$) is given by
\begin{eqnarray}
\relax [A_{1}\cdot A_{2}\cdot \dots \cdot A_n] &\equiv&
\frac{1}{n!}\sum_{perm.} (-1)^{\epsilon_{i_1\dots i_n}}
(A_{i_1}\cdot A_{i_2}\dots \cdot A_{i_n}), \label{antisym}
\end{eqnarray}
where $(A_1\cdot A_2\dots \cdot A_n)$ denotes the symmetric
product
\begin{eqnarray}
(A_1\cdot A_2 \cdot\dots  \cdot A_n) &\equiv& \frac{1}{2}(. ((A_1
A_2)A_3\dots )A_n) +\frac{1}{2} (A_1(A_2(\dots A_n)).).
\end{eqnarray}
In such a case one can show that the three-fold product of
octonionic $\Gamma$-matrices $\relax C[\Gamma_i\cdot \Gamma_j\cdot
\Gamma_k]$ provides $75$ antihermitian matrices, describing
together with $C$ an arbitrary $4\times 4$ octonionic
antihermitian matrix. The definition (\ref{antisym}), applied to
the five-fold products of octonionic $\Gamma$-matrices provides
their octonionic hermiticity. Further, by explicit calculation one
can show that there are $52$ independent real tensorial charges
describing the five-tensor sector of the octonionic M-algebra,
i.e. they span arbitrary $4\times 4$ octonionic-hermitian
matrices. We thus see that equivalently one can write the
octonionic M-algebra (\ref{newsusy}) as follows
\begin{eqnarray}
\relax \{Q_a,{Q_b}^\dagger\} &=& C_{ac} [\Gamma_{\mu_1}\cdot \dots
\cdot \Gamma_{\mu_5} ]_{cb} Z^{\mu_1\cdots \mu_5},
\end{eqnarray}
where out of the $462$ real antisymmetric $5$-tensorial charges of
the standard M-algebra only $52$ are linearly independent, due to
the relation
\begin{eqnarray}
\relax [\Gamma_{{\mu_1}\dots\mu_5}] &=&
{A_{[\mu_1\dots\mu_5]}}^\nu \Gamma_\nu +
{A_{[\mu_1\dots\mu_5]}}^{[\nu_1\nu_2]}\Gamma_{[\nu_1}\Gamma_{\nu_2]},
\label{M2M5}
\end{eqnarray}
with constant $c$-number coefficients
${A_{[\mu_1\dots\mu_5]}}^\nu$,
${A_{[\mu_1\dots\mu_5]}}^{[\nu_1\nu_2]}$ \footnote{The relation
(\ref{autom}) is a particular case of the formula (\ref{M2M5}).}.

The relation (\ref{M2M5}) implies that in $D=11$ octonionic
M-algebra there is an equivalence of the octonionic
five-superbrane and the octonionic two-superbrane (supermembrane)
sectors. We would like to stress again that\footnote{We note that
in order to express the octonionic structure as constraints on the
$528$ real Abelian tensorial charges describing the generalized
supersymmetry of standard M-theory we use the definition of
octonionic-valued anticommutator from footnote $6$.}
\par {\em a})
The octonionic supermembrane is characterized by constrained
number of two-tensorial charges - from $55$ to $41$. The remaining
$14$ generators of $G_2$ describe inner automorphisms of the
algebra (32) of octonionic units.
\par {\em b})
If we keep the $11$ degrees of freedom corresponding to the
momentum sector, the five-superbrane is also characterized by $41$
independent degrees of freedom i.e. $462$ degrees of freedom of
standard five-tensor charges  in $M$-theory are restricted very
much indeed. It is interesting to find a geometric interpretation
of such a huge reduction of degrees of freedom.

\par {\em c})
The $D=11$ Lorentz covariance algebra is described also by $41$
generators of $ \frac{O(10,1)}{G_2}$.

\par {\em d})
The anti-de-Sitter extension of octonionic $M$-algebra is
described by the octonionic extension of superalgebra $OSp(1;4)$,
which we denote by $U_{\alpha} U(4;1|{\bf O})$ (see also
\cite{HL}). The octonionic $M$-algebra is a contraction of
$U_\alpha U(4;1|{\bf O})$, when the anti-de-Sitter radius $R \to
\infty $.

\par
It should be pointed out that the octonionic M-algebra
(\ref{Moctalg}), whose superalgebraic structure for real
generators we computed with the prescription in footnote $6$, is a
Lie superalgebra.
\par
{\em ii}) {\em The octonionic conformal M-algebra.}

Let us recall that in the cases ${\bf F}={\bf R},{\bf C}$
 and ${\bf H}$ the ${\bf F}$-valued
supercharges $Q_{a}$ as well as the set of bosonic charges
$Z_{ab}$ were carrying the representation of the superalgebra
$U_{\alpha} (4;{\bf F})$, describing respectively $D=4$, $D=5$ and
$D=7$ standard anti-de-Sitter algebras. If we introduce the
algebra $U_{\alpha} (4;{\bf O})$, by means of $4\times 4$
octonionic-valued matrices $K_{ab}$, satisfying the relation
\begin{equation}\label{ll42}
  K^{\# }\, A = - A \, K
\end{equation}
where $A = - A^{T}$  (or in general $A^{\#} = -A$) describes the
antisymmetric (or in general case octonionic-antiHermitian)
metric, we obtain the $D=11$ de-Sitter-like  octonionic algebra
introduced in \cite{Ced}.

The octonionic conformal M-algebra $U_\alpha (8|{\bf O})$ which we
propose as defined by extending to octonions the scheme described
in (\ref{gradthree}) is graded as follows:
\begin{eqnarray}
&&\begin{array}{ccc}
 I_{-2}& I_0 & I_2\\
  {\tilde Z}_{ab}  & L_{ab}\subset gl(4|{\bf O})  & Z_{ab}
\end{array}
\end{eqnarray}
It contains $232$ real generators and
 we conjecture it has the following properties:

{\em a}) We postulate that using the generators of $U_{\alpha}
(8|{\bf O})$ one can obtain the realization of
$\frac{O(11,2)}{G_2}$, replacing  standard $D=11$ conformal
algebra.

{\em b}) The $128$ real generators of $Gl(4|{\bf O})$ describe the
$4\times 4$ octonionic matrix in the place of the general real
$Gl(32|{\bf R})$ covariance group of standard M-theory with $1024$
real generators.

{\em iii)} {\em The octonionic conformal $M$-superalgebra.}

The superextension of $U_{\alpha}U(8;1|{\bf F})$ to ${\bf F}={\bf
O}$ describes the octonionic conformal M-super-algebra, with
bosonic sector described by $U_{\alpha}(8|{\bf O})\times U(1|{\bf
O})$ (232+7=239 real generators), where the internal sector
$U(1|{\bf O}) \simeq S^7$ describes a parallelizable manifold
which only can be described by an extension of the notion of
standard Lie algebra - the so called soft Lie algebras
\cite{Soh2}. Some indications suggest that the structure of the
octonionic supergroups $U_\alpha U(4;1|{\bf O})$ and $U_\alpha
U(8;1|{\bf O})$ in the real basis is that of a (graded) Malcev
(super)algebra \cite{{CP},{CRT}}. We leave this investigation for
future work.

\section{Concluding remarks}

Our proposal is the extension of Kugo and Townsend \cite{KT}
relation between division algebras and standard sequence
$D=3,4,6,10$ of supersymmetries within the H\L S scheme to the
case of generalized supersymmetries in the dimensions $4,5,7$ and
$11$ . The idea that $d$-dimensional Minkowski space-time should
be extended by additional dimensions, describing tensorial central
charges coordinates, has been proposed already some time ago (see
e.g. \cite{{Fro},{ES},{Cur},{SR}}). Our framework provides a
concrete way of extending standard space-time framework to
dimensions $5$ and $7$ and ultimately to $D=11$.

In this paper we did not develop various aspects of the proposed
scheme. Let us only present a list of them, as problems for
possible further considerations.

{\em i}) One can ask if the choice of our sequence and its
space-time supersymmetry interpretation is unique. Indeed, because
$U_{\alpha}U(8;1|{\bf C})\equiv SU(4,4;1)$ include $D=6$ conformal
symmetries $O(6,2)$, one could also assign our sequences of
superalgebras to $D=4,6$ and $7$. The argument for using the
sequence $D=4,5,7$ comes from the link with minimal Clifford
algebra realizations (see (33)). The other choice of division
superalgebra sequence can be obtained if we replace the
quaternionic structure of $D=6$ Poincar\'{e} superalgebra by the
real one\footnote{In our case it is replaced by the complex one.}.
In such a case one obtains the sequence $U_{\alpha}U(16;1|{\bf
F})$ as describing generalized conformal superalgebras in $D=6$
($OSp(1;16|{\bf R})$), $D=7$ ($SU(8,8:1)$) and $D=9$
($U_{\alpha}U(16;1|{\bf H})$).

{\em ii}) For simplicity we do not consider here more explicitly
the extended generalized symmetries, but such a generalization is
obvious. In particular the extended generalized conformal
supersymmetry with $N$ copies of of ${\bf F}$-valued supercharges
${Q_a}^i$, ${S_a}^i$ is given by the superalgebra $U_\alpha U
(8;N|{\bf F})$, with the internal sector (R--symmetries) $U({\bf
F})$ ($O(N)$ for ${\bf F}={\bf R}$, $U(N)$ for ${\bf F}={\bf C}$
and $U(N;{\bf H})\equiv USp(2N)$ for ${\bf F}={\bf H}$).

 {\em iii}) Our considerations are on purely algebraic level.
One should also consider the representation theory  of the
generalized (super)symmetry algebras, e.g. express the generators
in terms of oscillators\footnote{In this respect see especially
\cite{Gun}).} and consider the complete set of Casimir's.

{\em iv}) We did not mention here generalized de-Sitter
supersymmetries
 and
mentioned only for ${\bf F}={\bf O}$ anti-de-Sitter symmetries and
supersymmetries. The generalized anti-de-Sitter superalgebras in
dimension $D$ should be identified with the generalized conformal
superalgebras in dimension $D-1$, i.e. our set of superalgebras
$U_\alpha U(8;1|{\bf F})$ describes the generalized anti-de-Sitter
algebras in $D=5,6,8$ and possibly $D=12$ (the last case for ${\bf
F}={\bf O}$). The discussion of generalized $D$-dimensional
de-Sitter superalgebras, equivalent to generalized Lorentz
superalgebras in dimension $D+1$
  is similar in principle, however
with differences in technical details (see also \cite{Ferr}).

{\em v}) We did not discuss here an important issue of
supersymmetric dynamics, covariant under generalized Poincar\'e
and conformal supersymmetries. We would like only to mention that
the preliminary results in such a direction has been already
presented \cite{{BL},{BLS},{ABL}}  for massless $D$0-superbranes
(supersymmetric particles) mainly in $D=4$ with $OSp(1;8)$ as
generalized conformal algebra.\\{\quad}
\\
 \noindent {{\bf Acknowledgments}}

The authors would like to thank Dmitri Sorokin for valuable
comments. One of the authors (J.L.) would like to thank  Jos\'e
Helayel-Neto for his hospitality at CBPF in Rio de Janeiro, where
the main part of this paper was prepared. F.T. is grateful for the
hospitality at the Institute of Theoretical Physics of the
University of Wroc\l aw, where this work has been completed.

\end{document}